\documentclass[prl,preprintnumbers]{revtex4}
\usepackage{graphicx,t1enc}
\usepackage{epsfig}
\usepackage{amssymb,amsmath}

\begin{document}

\title{Game theory in models of pedestrian room evacuation}

\author{S. Bouzat}
\author{M. N. Kuperman}

\affiliation{Consejo Nacional de Investigaciones Cient\'{\i}ficas y T\'ecnicas \\
FiEstIn, Centro At\'omico Bariloche (CNEA), (8400) Bariloche, R\'{\i}o Negro,
Argentina.}

\begin{abstract}
We analyze the pedestrian evacuation of a rectangular room with a single door considering a
Lattice Gas scheme with the addition of behavioral aspects of the pedestrians. The movement of the
individuals is based on random and rational choices and is affected by conflicts  between two or
more agents that want to advance to the same position. Such conflicts are solved according to
certain rules closely related to the concept of strategies in Game Theory, cooperation and
defection. We consider game rules analogous to those from the Prisoner's Dilemma and Stag Hunt
games, with payoffs associated to the probabilities of the individuals to advance to the selected
site. We find that, even when defecting is the rational choice for any agent, under certain
conditions, cooperators can take advantage from mutual cooperation and leave the room more rapidly
than defectors.

\end{abstract}

\maketitle

\section{Introduction}
The problem of enclosed pedestrian evacuation has been studied from various points of view.
On the one hand it presents academic interest because of its similarity with makeshift linked to
granular media \cite{helb}. On the other hand, understanding the dynamics of the movement of
pedestrians and anticipating the problems that may arise in an emergency situation is critical in
the design of large spaces that will be occupied by many people \cite{helb2}. The efficient
evacuation of the occupants of such places under a state of emergency is fundamental when trying to
minimize the negative effects of panic and  confusion, clogging and avalanches. Predicting
evacuation patterns is one of the first steps towards this goal. Evacuation is only a particular
aspect of a broader problem: the pedestrian movement.

Pedestrian dynamics  has been extensively studied from a theoretical
\cite{burs,taj,kir,hel1,hel3} and experimental point of view
\cite{mura,blue,taki,hel5,seyf}.

The associated pedestrian flow is usually modeled as a many-body system of
interacting individuals. The literature on this subject
is rather extensive \cite{hoog,baek,tani,zhen,yama,huan,migu,chra,bagl,fran}, exposing several
different approaches to the problem. In \cite{hel5}, the authors introduce the active walker model
to describe human trail formation and they show that the pedestrian
flow system exhibits various collective phenomena interpreted as
self-organized effects. In \cite{hend} the author suggest that the
behaviors of pedestrian crowds are similar to gases or fluids.
Other authors prefer the formalism of cellular automata to frame their models \cite{fuku}.
This is the approach we are going to adopt in the present work.

In the present Lattice Gas \cite{fri} model, $N$ pedestrians are set on the sites of a $L \times W$
lattice,
with the restriction that each site cannot hold more than one walker.
The pedestrians move to empty sites according to a preferential direction
dictated by the need of escaping from the room \cite{mura,mura2}. Conflicts
between agents that want to get to the same position are solved using
specific game rules and taking into account previously defined strategies (of cooperation or
defection)
which represent the characters of the agents. Hence, the considered pedestrian dynamics include
games between agents which affect their possibilities of leaving the room.

Real pedestrians are entities much more complex than automata with the sole idea of escaping from a
room.
One of the most difficult aspects of modeling pedestrian flow is to simulate the effects of
subjective
aspects that affect the interaction among the pedestrians \cite{seyf,seyf2,hel4,hao,hao2}.
Most of the models have focused on modeling the flow of individuals in pure mechanistic approaches
but
the behavioral reaction of the evacuees during their movement has not been so far investigated
in detail. By including a game dynamics in the resolution of conflicts between pedestrians, the
present work aims at opening a door to the analysis of this important and rather complex feature.

\section{2 x 2 symmetric games}

First, we want to introduce some basic concepts of game theory and 2 x 2 symmetric games that are
relevant for the pedestrian dynamics that we will consider.
A two players game  can be characterized by the set of the strategies that the players can adopt and
by
the payoff received by each strategy when confronting any other. If the game is symmetric, i.e. both
players
have access to the same set of strategies and payoffs, the information can be uploaded in an $n
\times n$ matrix, with $n$
the number of strategies. In 1966, M. Guyer and A. Rapoport \cite{rapo1} cataloged all the
$2 \times 2$ games. There are twelve symmetric games, eight of them are trivial, in the sense that
there is no conflict of interest as both players prefer the same outcome. The remaining four games
represent four distinct social dilemmas.

In a $2 \times 2$ game there are only two different strategies that can be defined as  cooperative
(C) and defective (D).

We can consider only relative payoff values, and this let us with four generic  payoffs represented in the following
matrix

\begin{center}
\begin{tabular}{|c|c|c|}

  \hline
  & C & D \\
  \hline

  C & $CC$ & $CD$ \\

  \hline

  D & $DC$ & $DD$ \\
   \hline
\end{tabular}
\end{center}

Here, each matrix element indicates the payoff of a player adopting the strategy at the file when competing with a
player adopting the strategy at the column.
$CC$ is the payoff to each of the two players, when both
cooperate; $DD$ is the payoff when both defect. When one player
cooperates and the other defects, the payoffs of the
cooperator and the defector are $CD$ and $DC$ respectively.

The relative values of the four payoffs characterize the whole family of symmetric $2 \times 2$
games.

In order to be consistent to the names assigned to the strategies, the payoff $CC$ must be preferred
to $CD$, meaning
that a cooperation from the opponent is always preferable to defection.
Also, $DC$ must be better than $DD$. Still, there is something missing to define a dilemma.
If defection is bad, it is natural to avoid it, unless there is
a temptation to defect. There are three situations in which this can occur.
Either there is an incentive to defect when the other player cooperates
($DC>CC$), or there is an incentive to defect when the other player
defects ($DD>CD$), or both.

There are four different games that fulfill at least one of the
aforementioned conditions.
\begin{itemize}
 \item Prisoner's Dilemma: $DC>CC>DD>CD$
 \item Stag Hunt: $CC>DC>DD>CD$
 \item Deadlock: $DC>DD>CC>CD$
 \item Chicken: $DC>CC>CD>DD$

\end{itemize}

We want to focus on the two first cases.
In particular, the Prisoner's Dilemma (PD) represents situations in which obtaining
cooperation is difficult because of the increased individual incentives to
defect. Despite that players can realize that they would be better if they both
cooperate than if they both defect, defecting is individually the best choice. It is a dominant
strategy. Under this scenario, organizational cooperation involves organizing the
individuals to work towards a common goal even if they have to give up personal incentives to
defect.

In turn, Stag-Hunt (SH) represents a situation in which coordination is difficult
because of the uncertainty about what the opponent will do.
Unlike in PD, both players realize that they are best off when they
coordinate on (C, C), but do not want to select C if
they think that the other player will not do so as well. Therefore,
organizational coordination involves be convinced  that others will work towards the common
goal as well, in which case it is individually rational for everyone to do so.

\section{Model}

We analyze the evacuation of a rectangular room of size $L\times W$ with a single door of length
$L_{d}$ located at the
center of one of the walls. We consider a discrete time and space dynamics in which the pedestrians
(or agents)
can occupy the sites of a square lattice an can perform jumps between first neighbor sites according
to certain rules.
When more than one agent attempts to jump to the same site at the same time a conflict occurs. To
solve the conflict
the involved agents (which can be between 2 and 4) play a game using previously defined strategies.
As a results of the
game, either there can be a winner which finally gets to the desired position, or there can be no
winner and all the involved
agents loose their opportunity to move at that time step.

The discrete sites are labeled as $(x,y)$ with $1\leq x\leq L$ and $1\leq y\leq W$. The sites with
$x=1, x=L$ and $y=W$ belong to the walls and cannot be occupied by agents. The sames occurs for the
sites with $y=1$ excepting for those
with $x_l\le x \le x_r$ that correspond to the door exit. We consider a symmetric position of the
door setting $x_l=L/2-L_d/2$ and $x_r=L/2+L_d/2$. For the sake of
simplicity, throughout the paper we consider $L_d=L/10$.

We consider an initial density of agents $0\le\rho\le 1$, corresponding to an initial number of
agents equal to $\rm{Int}(\rho\times L\times W)$. The agents are placed at random positions $1<x<L$,
$1<y<W$ at $t=0$.
Then, at each time steps, the dynamics involves three stages. First, every agent chooses a neighbor
site where to attempt to jump. Second, all the conflicts are identified and solved according to the
game rules. Finally, the winners of the conflicts (as well as
the agents that can move without conflict) jump to their desired positions. The agents that reach
the door
are taken out of the system. The simulation ends when all the agents have abandoned the room. In
the next paragraphs we explain each of the three instances in detail .

\subsection{First stage: Choosing the site where to attempt to jump}

Each agent can attempt to jump to any of the four neighbor sites which we simply label as up,
down, left and right (see figure 1.a).
With probability $R$ ($0\le R\le 1$), the direction of the jump attempt is chosen at random.
Conversely, with probability $1-R$ the direction of jump attempt is chosen according to a desired
(or rational) direction which is
defined for each agent pointing essentially toward the door in a way that we explain in detail
below.
Importantly, the angle $\alpha$ defining such desired direction is continuous between $0$ and
$2\pi$. Then the desired
direction is projected on the discrete allowed directions in order to define the probabilities for
attempt of jumps.
This is done as follows. For instance, consider the case depicted in figure 1.a. In such a case, we
define the probabilities
for attempt for a jump to the up, down, right and left directions as
$p_u=R/4,p_d=R/4 + (1-R) cos(\alpha)/Z, p_r=R/4$ and $p_l=\frac{1}{4}R + (1-R) \sin(\alpha)/Z$,
respectively.
Here, $Z=|\cos(\alpha)|+|\sin(\alpha)|$ is a normalization constant required to have
$p_u+p_d+p_r+p_l=1$.
Note that, in this case, $p_u$ and $p_r$ contain only the term proportional to $R$ describing the
random way of choosing,
while $p_d$ and $p_l$ include also a term which is proportional to $(1-R)$ and to the projections of
the desired direction. Now,
considering an arbitrary desired direction defined by an angle $0\le \alpha \le 2 \pi$ increasing
clockwise from $\alpha=0$ (corresponding to the down direction),
we have the general definition
\begin{eqnarray}
\label{pattempt}
p_u&=&\frac{1}{4}R + (1-R) \frac{(-\cos(\alpha))(1-{\rm sgn}(\cos(\alpha))}{2 Z} \nonumber \\
p_d&=&\frac{1}{4}R + (1-R) \frac{\cos(\alpha)(1+{\rm sgn}(\cos(\alpha))}{2 Z} \nonumber \\
p_r&=&\frac{1}{4}R + (1-R) \frac{(-\sin(\alpha))(1-{\rm sgn}(\sin(\alpha))}{2 Z} \nonumber \\
p_l&=&\frac{1}{4}R + (1-R) \frac{\sin(\alpha)(1+{\rm sgn}(\sin(\alpha))}{2 Z}.
\end{eqnarray}
Note that the desired or rational direction contributes only to the probabilities of jumping to the
discrete directions on which
it has a positive projection. Moreover, the probability $R$ measures the randomness of the motion in
such a way that
for $R=0$ only the two discrete direction defined by the desired direction are allowed, while for
$R=1$ all four directions
have equal probabilities.
\begin{figure}[h]
  \begin{center}
    \includegraphics[width=0.8\textwidth]{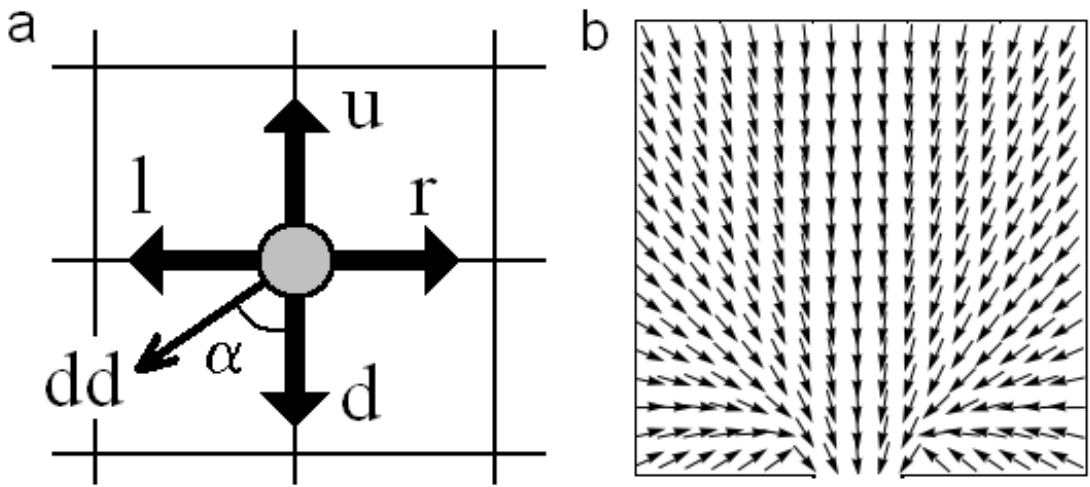}
    \caption{Model for agents motion. a) Allowed directions (up, down,
left, right) for the motion of an agent (central circle), and desired direction (dd). b) Field of desired
direction in a square room.}
\end{center}
\end{figure}

The desired direction is defined through a target position $(x_T,y_T)$ toward which the agents
point.
For an agent located at $x,y$ we set $x_T=(L+1)/2$ (independently of $x$ and $y$),
$y_T=-L/10+2L/5(-y/\sqrt{(y-L/2+5)^2+y^2}+\sin(\tan^{-1}(a)))$ for $y<a(x-x_{r})$ or $y<-a(x-x_{l})$
and $y_T=-L/10$
otherwise. Here, $x_r$ and $x_l$ are the right and left limits of the door and $a=3$. Figure 1.b
shows
the field of desired directions for a square room. The dependence of $y_T$ on $(x,y)$ is chosen in
order to produce
a recirculation pattern that prevents the agents to remain ''trapped'' close to the bottom wall.
This can be considered as equivalent to an effective repulsion exerted by the bottom wall. Such
recirculation pattern is important to get a realistic escape dynamics as the one shown in figure 2.
The consideration of a fixed target position (independent of the agent position)
produces non realistic dynamics close to the $y=1$ wall. For instance, for a fixed target position
at the center of the door, an agent located at the exit would walk along the door until it reaches
its center instead of getting out of the room immediately.

Now we can finally explain the detailed procedure for determining the attempt of jump for each agent
at each time step.
First, the desired direction and the probabilities given in equation (\ref{pattempt}) are computed.
Second,
using the probabilities (\ref{pattempt}), the agent performs two attempts of finding an empty
neighbor site for the jump.
This means that one of the four allowed directions is selected according to the probabilities
(\ref{pattempt}).
Then, in case the corresponding neighbor site is empty, it is marked as chosen for an attempt of
jump by the considered agent. Meanwhile, if the neighbor site is not empty, the procedure of
selecting one of the four
allowed directions according to the probabilities (\ref{pattempt}) is repeated once.
In case that the same or another non empty neighbor site is selected, the agent will not move at
that time step.

\subsection{Second stage: solving the conflicts}

When a given site is chosen by more than one agent on their attempt to jump, we say that
there is a conflict at that site. Thus, once all the pedestrian have chosen their sites where
to attempt to jump (excepting those that have lost their opportunity of jumping due to having
chosen a non empty site), all the conflicts have to be identified and solved in order to determine
the agents that will be able to move at the time step. The conflicts and their solutions imply thus
an effective interaction between the agents which affects the dynamics.

A conflict can involve two, three or four agents, due to the existence of four allowed directions
for arriving
to a given site. Each conflict will be solved through a game. As a result of the game, a {\em
winner} may
be selected to jump to the desired site, while the rest of the players (the {\em losers}) will lose
their opportunity to move at that time step. It is important to stress that, with certain
probability that we later indicate, a game can have no winner. In such a case, all the players will
lose their opportunity to move.

The definition of the game demands the consideration of strategies for the players.
We consider that from the beginning of the dynamics each pedestrian adopts an attitude that can be
either cooperative (C) or a defective (D). At the moment of the conflict, and according to the
strategies chosen by each of the individuals involved in the conflict, the players will be assigned
a probability to win, i. e. to jump to the desired site. In case of a competition between two
individuals the  considered probabilities are loaded into the following matrix, that is analogous to a
payoff matrix.
\begin{center}
\begin{tabular}{|c|c|c|}
  \hline
  & C & D \\
  \hline
   C & $\frac{1}{2}$ & 0 \\

  \hline

  D & $\frac{1}{P}$ & $\frac{1}{2P}$ \\
  \hline
\end{tabular}
\end{center}
Here $P\ge 1$ is a parameter that measures the conflictive attitude of the defectors,
which makes them focus more on the competition than on the possibilities of movement.
Thus, the matrix  summarizes our assumptions for the two players games, which are the following.
First, in a two cooperators game there is always a winner (that jumps to the desired position) which
is chosen at random between the two players. Second, in a game between a cooperator and a defector,
the cooperator has no chance to advance while the defector advance with probability $1/P$. Third, in
a game between two defectors, each of the agents has probability
$1/(2P)$ of becoming the winner. Note that the larger the value of $P$, the lower the probabilities
of stepping of the defectors, both when interacting with defectors and with cooperators. In
particular, in a game with at least
one defector, the existence of a winner is ensured only for $P=1$. Thus, values $P>1$ are used to
model situations in which defectors sometimes loose their opportunity to move due to their
conflictive attitude. In this sense, $P$ is a punishment to defection. Defection generates aversion
among cooperators, who when confronting a defector cooperate but with some resilience.
The interaction between defectors is competitive and non cooperative leading to a delay
in the solution of any conflict that may arise.

The two players game considered is straightforwardly analyzable in terms of game theory. We observe
that when $1\le P<2$ the matrix is analogous to the payoff matrix of the Prisoner's Dilemma, while
for values of $P$ higher than 2, the analogous situation is the Stag Hunt. As mentioned before, in
both cases there is an incentive to defect. While in the first case D is clearly a dominant
strategy, this is not true for the Stag Hunt. Still, in both cases, cooperators can only take
advantages from mutual cooperation, that is more attractive when $P>2$. The lower bound for $P$ is
imposed to fulfill the requirement $2 CC > DC + CD$ that prevents alternating cooperation and
defection in an iterative game.

The example can be generalized for encounters of $n$ players by defining the probabilities of winning
of a C and a D player in the cases in which all the rest are cooperators, and in the cases in
which the player (C or D) compete against $m<n-1$ cooperators and $n-1-m$ defectors. We consider such
probabilities defined by following matrix:

\begin{center}
\begin{tabular}{|c|c|c|}

  \hline
  & (n-1)C,\, 0\,D & m\,C,\, (n-1-m)D \\
  \hline
  C & $\frac{1}{n}$ & 0 \\
  \hline
  D & $\frac{1}{P}$ & $\frac{1}{(n-m)^2 P}$ \\
  \hline
\end{tabular}
\end{center}

The contents of the matrix can be summarized as follows. In a game between $n$ cooperators and
no defectors, there is always a winner. One of the players is randomly chosen to move
(with probability equal to $1/n$). In a game with at least one defector, the existence of a winner
is not ensured. In this case, the cooperators always loose and and each defector has probability
$\frac{1}{(n-m)^2 P}$ of winning, with $m$ the number of cooperators in the game. The probability of
having a winner is $\frac{1}{(n-m) P}$.

\subsection{Third stage: movement of agents}

As stated before, once all the conflicts have been solved, all the winners as well as all the agents
that can move without conflict are finally moved to their selected sites. In case that one agent
gets to a position with $y=1$ (i.e. it reaches the exit), it is taken out of the system and the
number of escaped agents is increased by one.

\section{Results}

We  characterize the evacuation dynamics of the system by computing the {\em mean exit time} which
is
defined as the average over realizations of the number of time steps required to evacuate completely
the room.
We  also study the time dependence of the {\em mean number of escaped agents}, i.e. the average over
realizations
of the number of agents that have reach the exit at a given time.

Before considering the general case of a population with both cooperators and defectors, we study
the
extreme cases in which all the agents have the same strategy.

\subsection{Only Cooperators: the random game case}

First we consider a population in which all the agents are cooperators. In this case, any conflict
leads to a random game in which there is always a winner. Namely, for a game between $n$
cooperators, each of them has a probability $1/n$ of becoming the winner and, thus, moving to the
desired site. Note that the parameter $P$ results irrelevant. In Figure 2 we show the positions of
all the agents at six instances of the evolution for a single realization of the dynamics for a
system of $L=W=50$.

\begin{figure}[h]
  \begin{center}
    \includegraphics[width=0.8\textwidth]{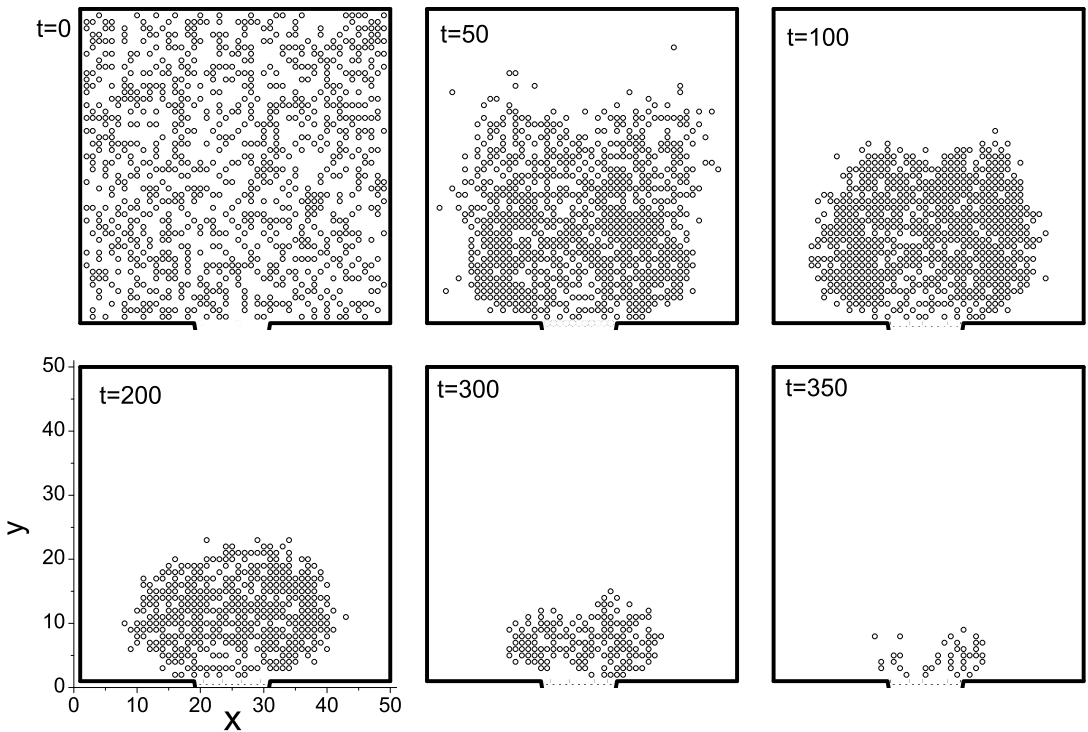}
    \caption{Room evacuation for a population with only cooperators. Parameters $L=W=50, \rho=0.4,
R=0.3$.}
  \end{center}
\end{figure}

\begin{figure}[h]
  \begin{center}
    \includegraphics[width=0.8\textwidth]{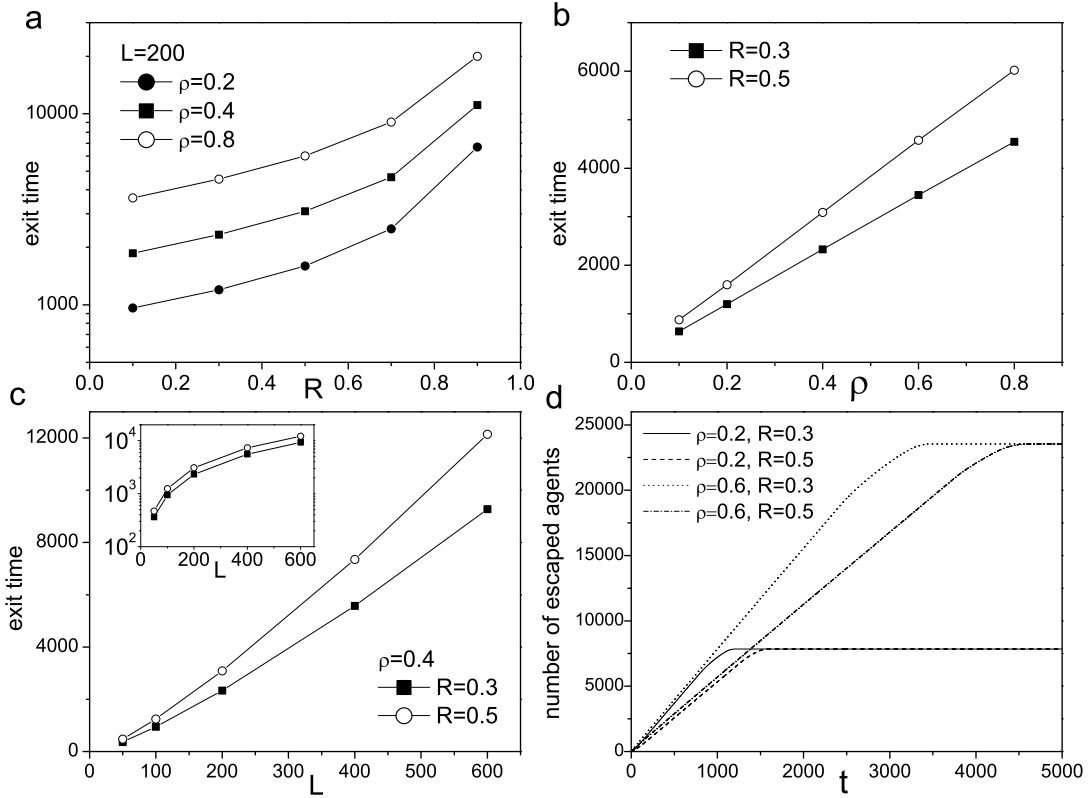}
    \caption{\label{figrandom} Results for a population of only cooperators in a square room
($L=W$). a) Exit time as a function of $R$ for different values of $\rho$ and $L=200$. b) Exit time
as a function of $\rho$ for different values of $R$ and $L=200$. c) Exit time as a function of $L$
for different values of $\rho$. The small inset shows the same curves in logarithmic scale. d)
Evolution of the number of escaped agents for different values of $\rho$ and $R$ for $L=200$.}
  \end{center}
\end{figure}

Figure (\ref{figrandom}) shows results for the mean exit times and the number of escaped agents as a
function
of time for the evacuation dynamics from a square room considering different system parameters
$\rho, R$ and $L$.
In Fig.(\ref{figrandom}).a we see that the mean exit time increases with $R$ more rapidly than
exponentially.
This means that the randomness of the dynamics strongly slows the evacuation process. Note that the
randomness
may be associated to an uncertainty of the agents in their knowledge of the position of the door. We
can also see that the mean exit time increases with the initial density, as could be expected.
Figure (\ref{figrandom}).b shows that such a growth
with $\rho$ is linear at fixed $R$.

In Fig.(\ref{figrandom}).c we study the dependence of the mean exit time on the system size. We see
that the growth
is slower than exponential and faster than linear.

Finally, Fig.(\ref{figrandom}).d shows the evolution of the mean number of escaped agents for
different values of
$\rho$ and $R$. Here, we see that the growth is linear along most of the evolution. Small deviations
from the linear
regime occur for very short times and very long times. This is due to the fact that the flux of
pedestrian out of the room is controlled mainly by the local density at the exit (and by the
parameter $R$ which is constant). The local density at the exit at small times coincides with
$\rho$, then it increases until it reaches a quasi-stationary value. Such quasi-stationary density
determines the slope of the linear growth of the evacuation profile. Finally, at large times, when
the evacuation process is about to finish, the density at the exit decreases and the flux at the
door is reduced. Fig.(\ref{figrandom}).d also shows that the slope of the linear growth is
essentially independent of the initial density. This indicate that the quasi-stationary value of the
density at the door depends only on $R$, as could be expected.

\subsection{Only Defectors}

Now we consider a population in which all the agents are defectors. This is an extreme case
opposite to the one considered in the previous subsection. We focus on the dependence of the results
on the parameter $P$ which now becomes relevant as it rules the probabilities of motion resulting
from all the conflicts.

\begin{figure}[h]
  \begin{center}
    \includegraphics[width=0.8\textwidth]{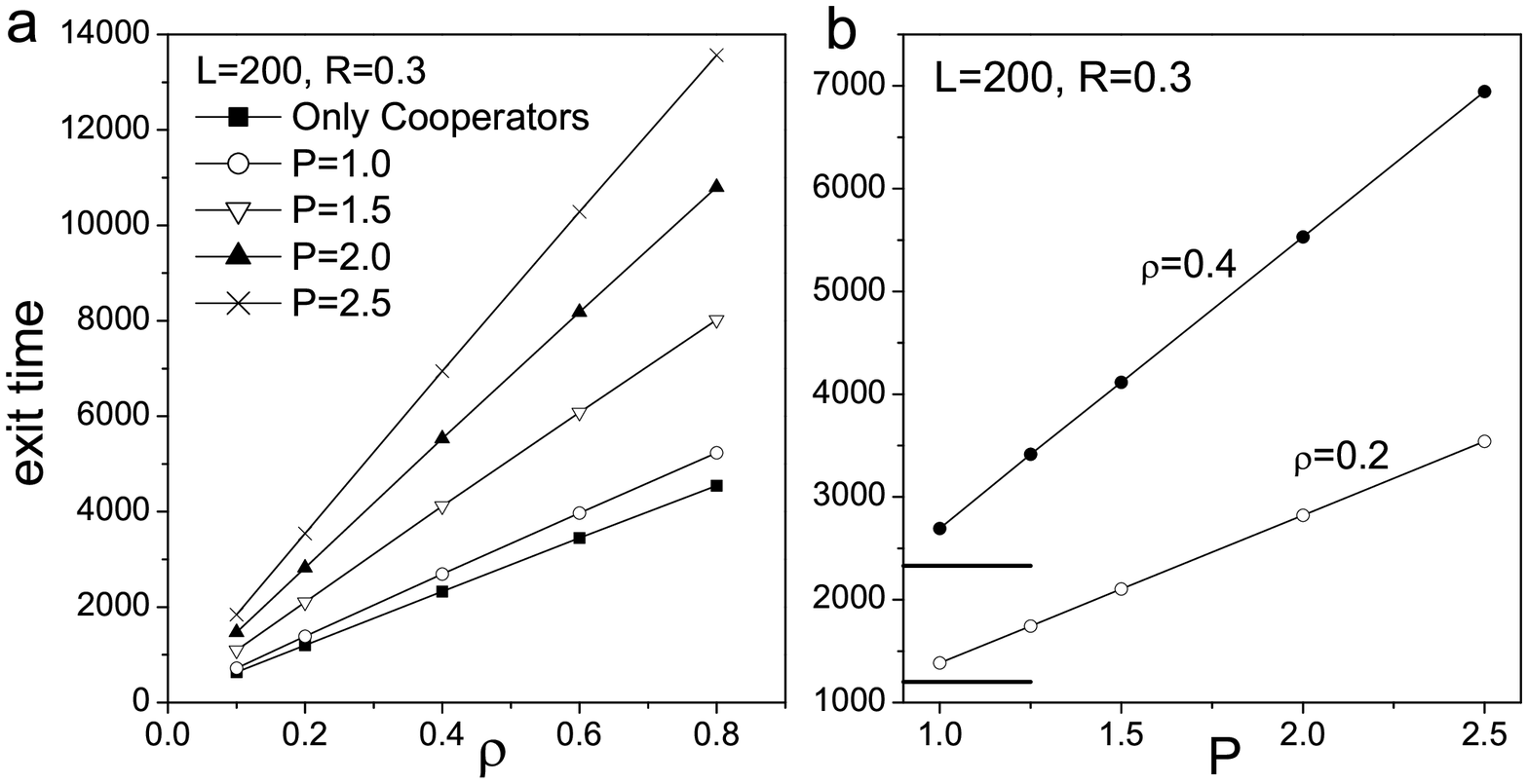}
    \caption{\label{figallD} Evacuation dynamics for systems with only defectors. a) Mean exit time
as a function
of the initial density of agents $\rho$ for different values of $P$. For the sake of comparison we
also indicate
the results for a system with only cooperators (squares). b) Mean exit time as a function of $P$ for
different
values of $\rho$. The segments on the left indicate the values for systems with only cooperators.}
  \end{center}
\end{figure}

Figure (\ref{figallD}).a shows the mean exit time as a function of the initial density $\rho$ for
different values
of $P$, while Figure (\ref{figallD}).b shows the mean exit time as a function of $P$ for different
values of $\rho$.
For the sake of comparison, in both insets we also include the results for the only-cooperators case
studied in the previous subsection. It can be seen that the mean exit time grows linearly both with
$\rho$ (Fig.(\ref{figallD}).a) and with $P$ (Fig.(\ref{figallD}).b). Moreover the mean exit time for
the only-collaborators case is always smaller
than that for the only-defector case at any value of $P$. This last fact can be understood as a
consequence of the delays for stepping after the conflicts, which are present only for games with defectors.

\subsection{Heterogeneous Populations}

Now we study the evacuation problem for heterogeneous populations with both collaborators and
defectors.

In Figure 5.a we show the mean exit time as a function of the initial fraction of defectors $\rho_D$
for fixed values of $\rho, R$ and $L$ considering different values of $P$. It can be seen that for
$P=1$ the exit time is almost independently of $\rho_D$. This is so because in this case the penalization
for defectors is null excepting for games with more than two defectors, which are likely only at large
$\rho_D$. At larger values of $P$ the exit time grows approximately linearly with $\rho_D$. In Figure 5.b
we show the normalized exit time, defined as $(t(\rho_D)-t(\rho_D=0))/t(\rho_D=1)$, where $t(\rho_D)$
is the mean exit time obtained for an initial fraction of defectors $\rho_D$. The plot shows us that the curves
for large enough $P$ are close to collapse, indicating thus the approximately linear behavior with $\rho_D$. However
a close look to the figure reveals that the dependence on $\rho_D$ may be exactly linear only for a value
of $P$ close to $P=2$ (i.e. the limit between PD and SH games), while it is superlinear for smaller values of $P$
(PD game) and slightly sublinear for larger values of $P$ (SH game).

\begin{figure}[h]
  \begin{center}
    \includegraphics[width=0.8\textwidth]{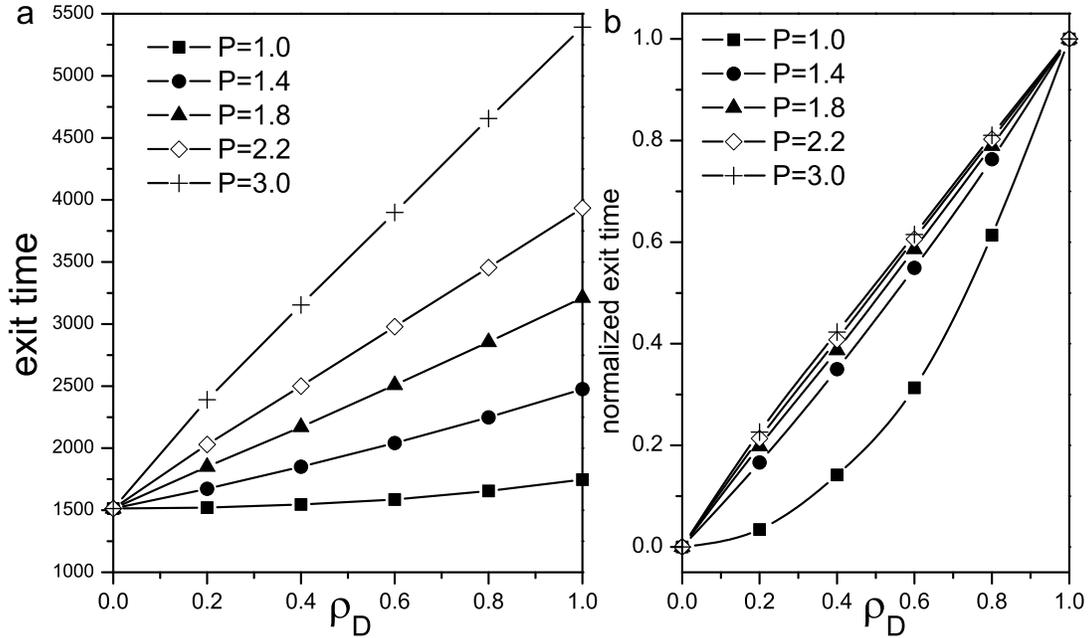}
    \caption{\label{figallM} Evacuation dynamics for heterogeneous populations. a) Exit time as a
function of the
initial fraction of defectors for different values of $P$. b) Normalized exit time as a function of
the
fraction of defectors for the same systems as in (a). All the calculations are for $L=200, \rho=0.4$
and $R=0.3$.}
  \end{center}
\end{figure}

\begin{figure}[h]
\begin{center}
    \includegraphics[width=0.8\textwidth]{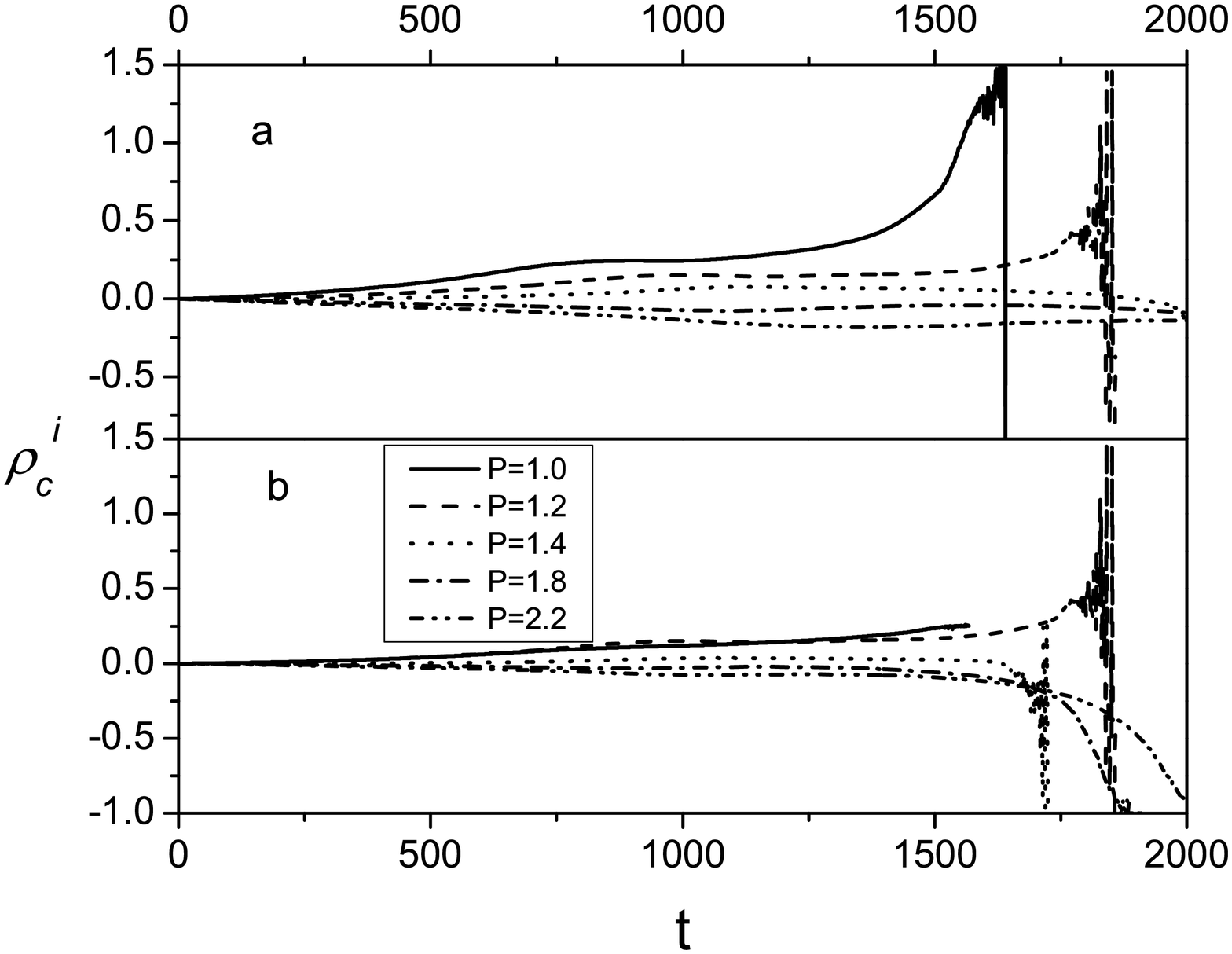}
    \caption{\label{room} Normalized instantaneous fraction of cooperators within the room for
different values of $P$. Results for $L=W=200$, $R=0.3$ and $\rho=0.4$ considering an initial
fraction of cooperators equal to $\rho_c(0) =0.4$ (a) and $\rho_c(0)=0.8$ (b).}
  \end{center}
\end{figure}

One important questions we pose is whether cooperators can or cannot take advantages from mutual
cooperation as it was verified in other systems where cooperation arise as an emerging phenomena
\cite{doe,egui,fu,gg,lang,lei,sza2,kup1,kup2,roc}.
With this goal in mind we characterize the dynamics of the system by three quantities, all of them
aiming at revealing the relative success of the cooperators in finding the exit.
First, we sample the composition of the population inside the room,  comparing the instantaneous
fraction of cooperators with the initial one.
We define a normalized instantaneous fraction of cooperators in the room as
\begin{equation}
\rho_c^i(t)=\frac{\rho_c(t)-\rho_c(0)}{\rho_c(0)},
\end{equation}
where $\rho_c(t)$ is the fraction of cooperators in the room at time $t$.
 A departure of the derivative of  $\rho_c^i(t)$ from zero indicates that the leaving individuals do
not represent
a random sample of the population in the room. A positive (negative) value of $\rho_c^i(t)$
indicates that the population in the room has a greater (lower) fraction of cooperators than
the initial. Meanwhile, a positive (negative) derivative of $\rho_c^i(t)$ indicates that defectors
(cooperators) are being more successful in finding the exit. Figure \ref{room} show the time
behavior of this quantity for two values of the initial density
of defectors and several values of $P$ for two different values of $\rho$. We observe that for small
values of $P$ we have $\rho_c^i>0$ throughout the evolution meaning that the defectors clearly
surpass the cooperators in finding the exit. Note that, not only $\rho_c^i(t)$ is positive but its
derivative increases with time, indicating the
continuous increment of the fraction of cooperators in the room. But as $P$ increases and even
still in the PD regime ($P=1.8$ in the figure), the values of $\rho_c^i$ become negative meaning
that the cooperators start to profit from mutual cooperation. While defectors loose time in futile
arguments, the cooperators leave the room. $\rho_c^i(t)$ and its derivative are negative, reflecting
the ability of cooperators to reach the
exit.
We have studied several initial configurations with different initial conditions including varying
values for the initial density of individuals, the initial fraction of cooperators and defectors and
several sizes.
With some small variations, we have found qualitatively the same results indicating the prevalence of
defectors at small $P$ and the prevalence of collaborators at large $P$.

\begin{figure}[h]
  \begin{center}
    \includegraphics[width=0.8\textwidth]{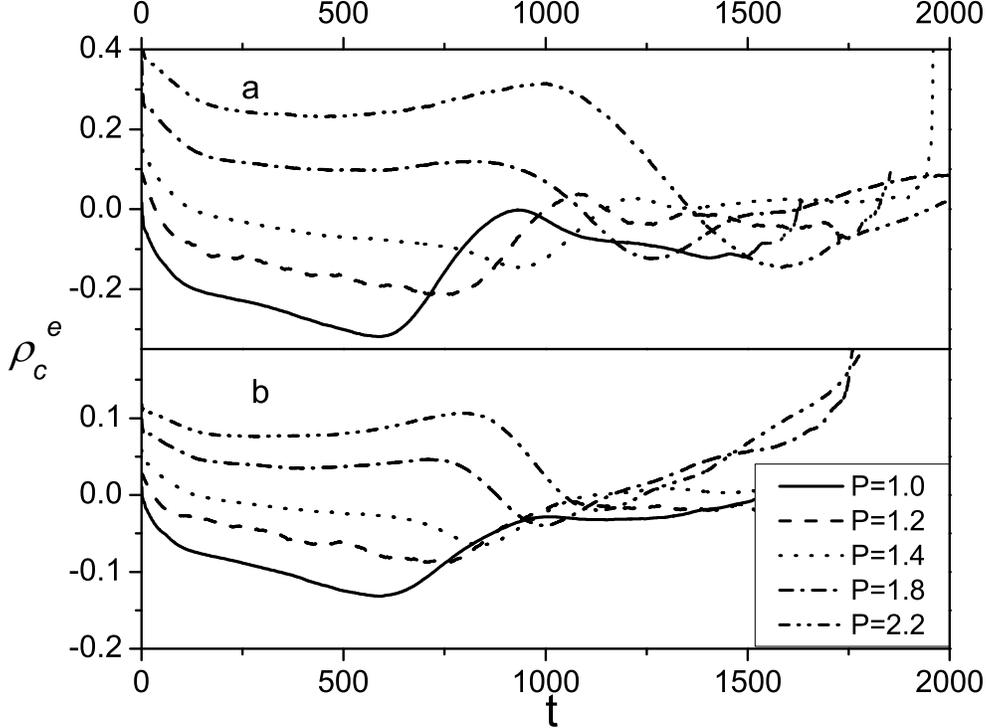}
    \caption{\label{exit} Normalized fraction of cooperators exiting the room for different values
of $P$. Results for $L=W=200$, $R=0.3$ and $\rho=0.4$ considering an initial fraction of cooperators
equal to $\rho_c(0) =0.4$ (a) and $\rho_c(0)=0.8$ (b).}
  \end{center}
\end{figure}

To complement this measure, we analyze the strategy of the individuals exiting the room at
each time step and calculate the fraction of cooperators among them. We compare this fraction
with the corresponding to the cooperators remaining in the room and define the normalized fraction
of exiting cooperators as
\begin{equation}
\rho_c^e(t)=\frac{\eta_c(t)-\rho_c(t)}{\rho_c(t)}.
\end{equation} Here $\eta_c(t)$ is the ensemble averaged fraction of cooperators among exiting
individuals
at time $t$. Figures \ref{exit} show the time behavior of $\rho_c^e(t)$ for the same conditions
situations analyzed in figure \ref{room}.
Again, the fraction of exiting defectors is higher than the expected one for small values of $P$ but
the situation is reversed as $P$ increases.

As cooperators can only advantage defectors when interacting only among them, any evidence of
cooperators performing
better than defectors at leaving the room must be reflected in the formation of clusters of
cooperators where mutual
cooperation occurs. If cooperators are not clustered, the defectors will outstrip them. The effect
of clustering of cooperators can be visually appreciated, but the assertion could be subjective.
So we define a quantity that let us measure the instantaneous degree of clustering of cooperators by
counting the
fraction cooperating neighbors of each cooperating individual $i$ in the room $\omega_c(i,t)$ and
then we define
\begin{equation}
C_c(t)=\frac{1}{N_c(t)}\sum \omega_c(i,t)/\rho_c(t),
\end{equation}
where $N_c(t)$ and $\rho_c(t)$ are the number and the fraction of cooperators in the room at time
$t$.
We exclude isolated individuals from this count.
\begin{figure}[h]
  \begin{center}
    \includegraphics[width=0.8\textwidth]{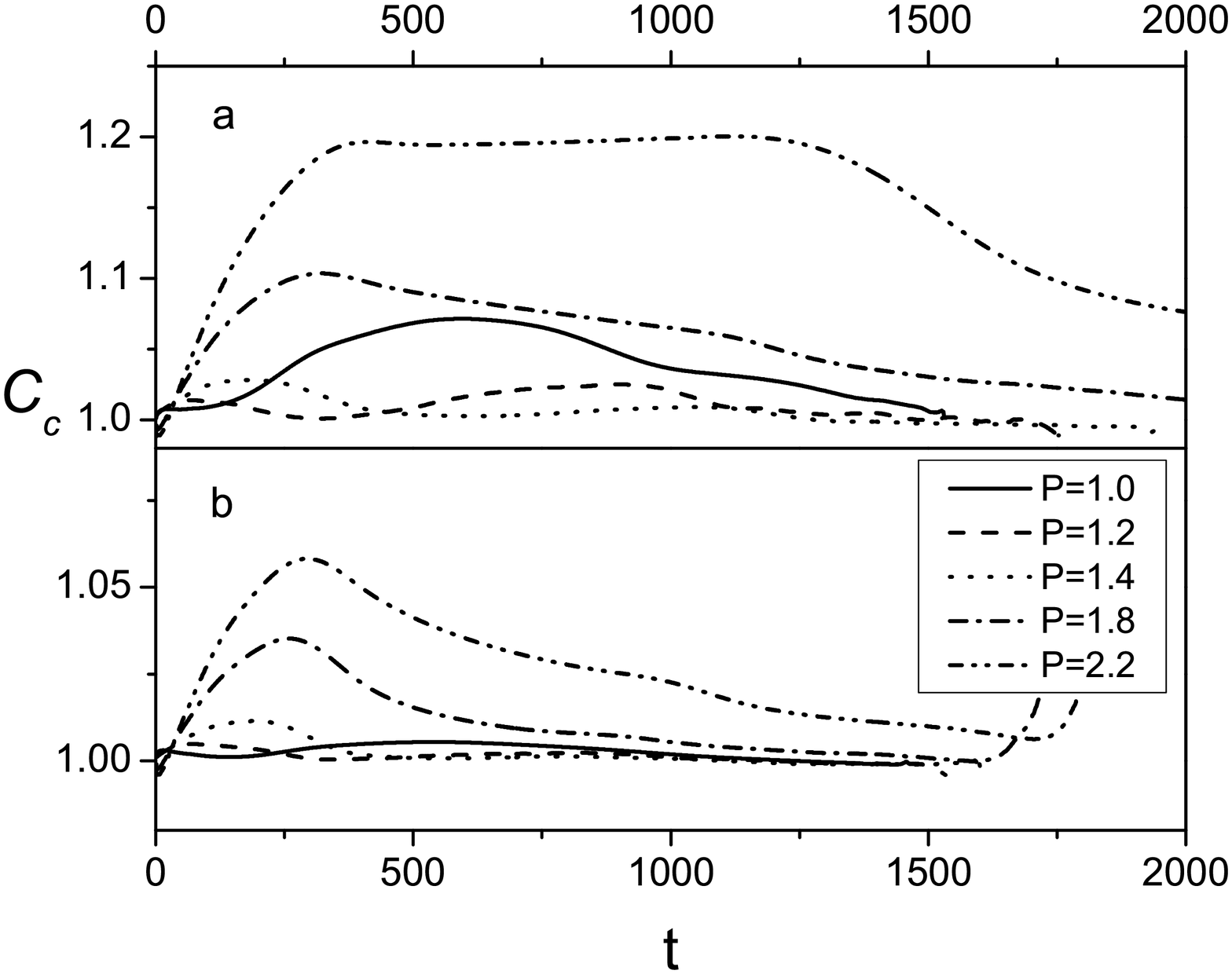}
    \caption{\label{clust} Clustering of cooperators in the room for different values of $P$.
Results
for $L=W=200$, $R=0.3$ and $\rho=0.4$ considering an initial fraction of cooperators equal
to $\rho_c(0) =0.4$ (a) and $\rho_c(0)=0.8$ (b).}
  \end{center}
\end{figure}

The results in figure \ref{clust} confirm the occurrence of the expected behavior. When the
quantities $\rho_c^i(t)$ and $\rho_c^e(t)$ indicate that cooperators are performing better than
defectors in reaching the exit, the results for $C_c(t)$ indicate an increase of the clustering of
cooperators. The dynamics of the system leads to a partial segregation into cluster of cooperators
and defectors, that leave the former in a situation of taking advantage from the benefits of mutual
cooperation. Figure \ref{clust} shows also an interesting behavior for low values of $P$. The curves
show an increase of the clustering of cooperator, more apparent for $P=1$ in Fig 1.a. This effect is
different in nature and shape from those observed for higher values of $P$. The origin of this increment
correspond to the fact that defectors, freely to move due to low values of punishment manage to approach
the exit, displacing the cooperators and producing a segregation that isolates the last from the door. This
effect is verified in Figs. \ref{room} and \ref{exit}, where we can observe that cooperator have
difficulties in  reaching the exit for values of $P$ close to 1.
This effect is also reflected in the fact that when the fraction of defectors is higher (0.6 in Fig.
\ref{clust}.a vs. 0.2 in Fig \ref{clust}.b), the displacement of cooperators from the door and their
corresponding segregation is enhanced.

The results thus indicate that cooperators can take advantage from mutual cooperation for
large enough values of $P$, typically $P>1.5$, well within the PD regime where still the 
DC payoff is larger than the CC one. The benefits from mutual cooperation are further enhanced
in the SH regime ($P>2$).

\section{Conclusions}

One of the shortcomings of the Lattice Gas models used in pedestrian dynamics is the lack of
inclusion of behavioral aspects. Considering this situation, the present work has a clear goal:
To develop a simple model able to consider some individual attitudes that may affect the movement of
interacting pedestrians. Motivated by this objective, we have amalgamated models for pedestrian
evacuation with Game Theory concepts and analyzed the emergence of non trivial effects that may
manifest.
The results show that the emerging phenomena observed in an Evolutionary Prisoner's  Dilemma
are also present here. In these works \cite{doe,egui,fu,gg,lang,lei,sza2,kup1,kup2,roc} the systems
are spatially extended and allow for the cooperators to form clusters capable to resist the invasion
of defectors and even to expand. In the present work the evolutionary aspects has not been so far included
but the growth of clusters is promoted by the mobility of the individuals. As a result of this
clustering, the cooperators can profit from the only advantage they have over the defectors: the
mutual cooperation. When this happens, cooperators have more success in reaching the exit than
defectors, as reflected in the plots.
It is important to understand the limitation and scope of the present
model. We are not intending to reproduce a real situation but to point out the effects that
behavioral aspects may have on the denouement of an emergency evacuation scenario.
Future work envisions the inclusion of evolutionary strategies, differentiated social roles, off
lattice dynamics and different geometries and obstacles.

\end{document}